\documentclass[12pt]{iopart}
\usepackage{iopams}
\usepackage{graphicx}
\usepackage{bm}
\usepackage{cite}
\usepackage{caption}

\begin{document}
\title[Efficient tracking of myosin II filament position and orientation over time]{A Python based automated tracking routine for myosin II filaments}

\author{L.S. Mosby$^{1, 2}$, M. Polin$^{1, 2}$, D.V. K\"oster$^1$}

\address{$^1$ Centre for Mechanochemical Cell Biology, University of Warwick, Coventry CV4 7AL, UK}
\address{$^2$ Physics Department, University of Warwick, Coventry CV4 7AL, UK}
\ead{d.koester@warwick.ac.uk, m.polin@warwick.ac.uk}

\begin{abstract}
The study of motor protein dynamics within cytoskeletal networks is of high interest to physicists and biologists to understand how the dynamics and properties of individual motors lead to cooperative effects and control of overall network behavior. Here, we report a method to detect and track muscular myosin II filaments within an actin network tethered to supported lipid bilayers. Based on the characteristic shape of myosin II filaments, this automated tracking routine allowed us to follow the position and orientation of myosin II filaments over time, and to reliably classify their dynamics into segments of diffusive and processive motion based on the analysis of displacements and angular changes between time steps. This automated, high throughput method will allow scientists to efficiently analyze motor dynamics in different conditions, and will grant access to more detailed information than provided by common tracking methods, without any need for time consuming manual tracking or generation of kymographs.
\end{abstract}

single particle tracking, automated detection, myosin

\submitto{\JPD}

\maketitle

\section{Introduction}
Molecular motors are important for many cellular processes such as cell cortex dynamics, cell migration, and the intracellular transport of vesicles. Purification of various molecular motors from tissue samples or after recombinant expression, and the analysis of their biochemical and bio-physical properties in reconstituted systems, was instrumental in furthering our understanding of the motors and how they are regulated in live cells. Whereas most molecular motors operate individually or in dimers, myosin II motors usually form bundles of about $10$ (non-muscular myosin II) up to $250$ (muscular myosin II) proteins at physiological salt conditions, giving rise to cooperative effects between the myosin head domains that govern the effective binding dynamics, speed and processivity of these myosin II protein ensembles. After these cooperative effects were studied theoretically, recent advances in microscopy and the design of reconstituted acto-myosin networks have allowed the study of the dynamics of myosin II filaments experimentally \cite{Melli2018, Mosby2019}. Traditionally, motor dynamics were studied using kymograph analysis or common single particle tracking (SPT) routines that identify point-like structures. These methods have drawbacks, for example kymographs only track dynamics along a given path and thus cannot be used to study two-dimensional diffusive motion. Similarly, common SPT routines cannot take into account the myosin filament orientation, which is a valuable parameter that can be used to characterise their dynamics. Here, we applied an image analysis routine from Astrophysics that was designed to identify galaxies \cite{Bertin1996,Barbary2016,Barbary2018} to detect myosin II filaments, as both types of object are characterised by their elongated, elliptical shapes with limited signal to noise ratios in large, heterogeneous samples. After particle detection, tracks are generated by comparing the positions, orientations and detected areas of particles between subsequent frames. This Python based routine worked robustly in the detection of myosin II filaments in a large, crowded sample set and made it possible to distinguish between the diffusive and processive motion of the filaments.     

\section{Methods}

\subsection{Experimental Data}
The experimental data used here to demonstrate the detection and tracking of myosin II filaments was obtained by interferometric scattering microscopy (iSCAT), as described previously \cite{Mosby2019, Young2018}. Actin filaments and skeletal myosin II filaments were purified from chicken breast muscle following established protocols. Glass coverslips ($\#1.5$ borosilicate, Menzel, Germany) were cleaned in a sequence of $2\%$ Hellmanex (Hellma Analytics, Mühlheim, Germany) followed by thorough rinses with EtOH and MilliQ water, and were blow dried with N\textsubscript{2} before being used for the preparation of experimental chambers. After formation of supported lipid bilayers (containing $98\%$ DOPC and $2\%$ DGS-NTA(Ni\textsuperscript{2+}) lipids (Avanti Polar Lipids Inc., US)) and addition of our actin-membrane linker protein HKE (decahistidine-ezrin actin binding domain), polymerized actin filaments and myosin II filaments were incubated in KMEH ($50 mM KCl, 2 mM MgCl_{2}, 1 mM EGTA, 20 mM HEPES, pH 7.2$) to allow the formation of membrane tethered acto-myosin networks \cite{Koster2014, Mosby2019}. 
Images were recorded on an iSCAT microscope setup equipped with a cMOS camera and processed to reflect interferometric contrast values in 32bit with a pixel size of $0.034\, \textnormal{x}\, 0.034\, \mu \textnormal{m}^2$ \cite{Mosby2019}.

\subsection{Myosin II Filament Detection \label{sec:Detection}}

We used the Python programming language as a platform for the myosin II filament detection based on the Python library for Source Extraction and Photometry (SEP) \cite{Bertin1996,Barbary2016,Barbary2018}, which generates the position, spatial extent, and orientation of ellipsoidal particles for each frame of a time lapse image sequence. 

To distinguish individual myosin II filaments, we first applied a signal and area threshold to isolate probable myosin II filament detection from background, followed by the application of a refined area threshold to distinguish individual myosin II filaments from aggregates and to reduce the rate of false-positive detection due to poor signal to noise ratios. For the first step, a detection was defined as a region of $>40$ pixels, each with an intensity of at least $1.5\sigma_{\textnormal{g,rms}}$ above the local background intensity, where $\sigma_{\textnormal{g,rms}}$ is the global root-mean-square error of the spatially varying background of the image. These minimum area and intensity cutoffs allowed confident detection of myosin II filaments without falsely detecting background fluctuations.

\begin{figure}[t!]
	\includegraphics[width=1.0\linewidth]{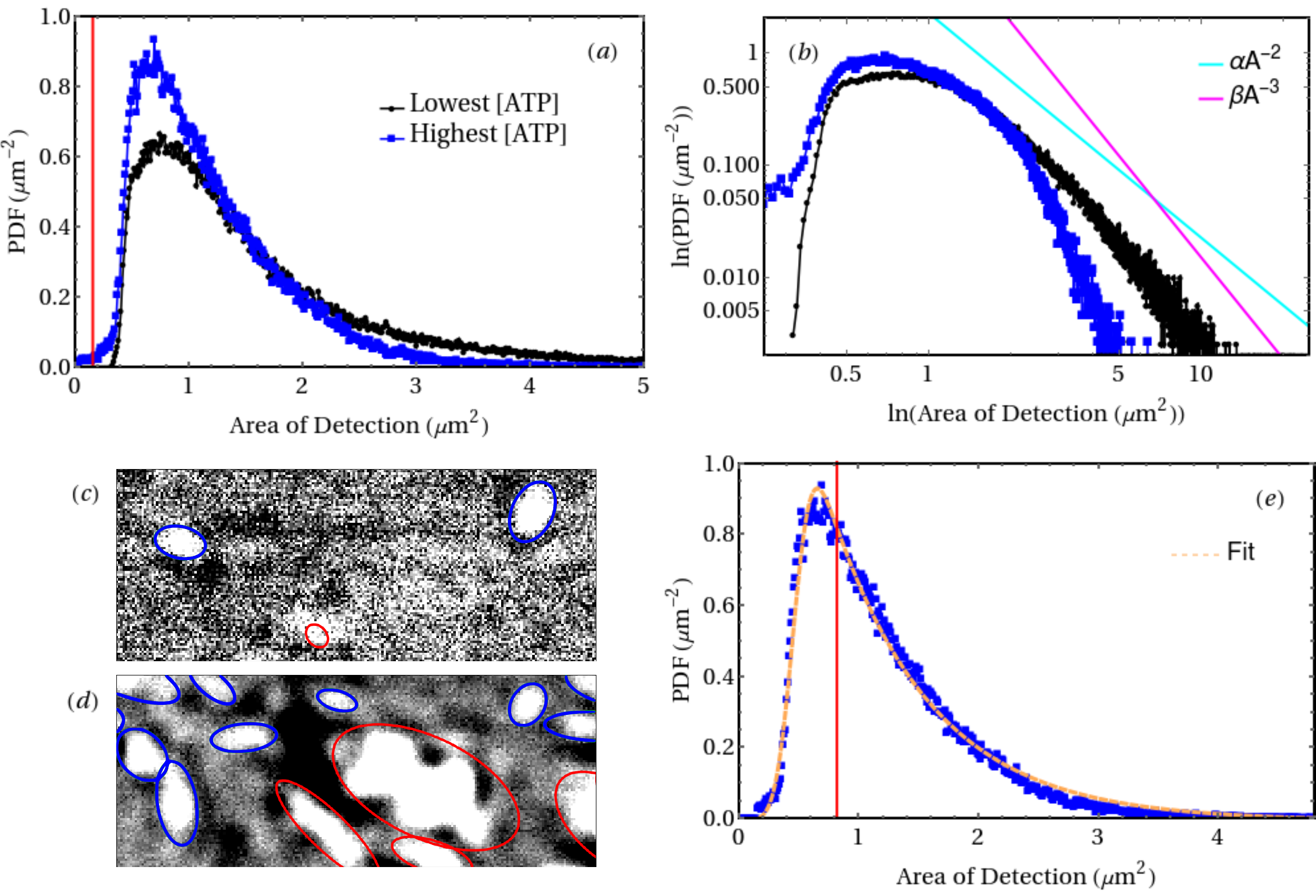}
	\caption{(a) Probability distribution functions of detection areas for the sample data sets with the lowest (black) and highest (blue) ATP concentrations. Decreasing the ATP concentration increases the rate of collisions between myosin II filaments, leading to aggregation and a higher proportion of the population in the exponentially decaying tail of the distribution. Red vertical line shows the minimum area cutoff $A_{\textnormal{min}}$, equal to the minimum area detection in the high signal-to-noise ratio data set with the lowest ATP concentration. (b) Log-log plots of the distributions showing their decay profiles compared to lines of constant gradient in log-log space. (c) Example of anomalously small area detection in the data set with the highest ATP concentration (red) due to poor signal to noise ratio and a disperse region of pixels with intensity greater than $1.5\sigma_{\textnormal{g,rms}}$ above the local background intensity. (d) Examples of myosin II filament aggregates in the data set with the lowest ATP concentration (red). (e) The corrected distribution after the removal of anomalously small area detection from the data set with the highest ATP concentration. An exponentially-modified Gaussian distribution was fitted to the data (orange), and the red vertical line shows the average area $A_{\textnormal{av}}=1.25A_{\textnormal{mode}}$.}
	\label{fig:AreaPanel}
\end{figure}

The SEP Python package computes a set of ellipse parameters (a: semi-major axis, b: semi-minor axis, $\theta$: orientation on a $-\pi/2 \leq \theta \leq \pi/2$ domain) for each detection based on the spatial dispersion of its intensity profile (sections 10.1.5 and 6 of v2.13 of the S Extractor User's Manual by E. Bertin). The area of each detection was then calculated as $A = \pi a b$ (figure \ref{fig:AreaPanel}(a,b)). For clarity the ellipse parameters were scaled to $x = 6x_{\textnormal{detected}}$ ($x = a \textnormal{ or } b$) to allow for better representation of the detected object by eye \cite{Barbary2018}. It should also be noted that the observed signal is a result of the convolution between the real signal and the point-spread function of the microscope. For these reasons the areas calculated using this method have only been used for diagnostics.

In order to prevent the false-positive detection of background intensity fluctuations (figure \ref{fig:AreaPanel}(c)) a minimum area cutoff was implemented. After considering the probability distribution function of the detected myosin II filament areas for a sample data set with a high signal to noise ratio, the minimum area cutoff was defined as equal to the smallest detected area $A_{\textnormal{min}} = 0.151\,\mu\textnormal{m}^2$. An average area for a single myosin II filament detection was also calculated as $A_{\textnormal{av}}=1.25A_{\textnormal{mode}}$, where $A_{\textnormal{mode}}$ was the maximum value of an exponentially-modified Gaussian distribution fitted to the area probability distribution function of a data set recorded at high ATP concentration displaying minimal myosin II filament aggregate formation (figure \ref{fig:AreaPanel}(e)). At lower ATP concentrations the sample data sets contained many myosin II filament aggregates (figure \ref{fig:AreaPanel}(d)), which resulted in a higher than expected rate of detection of large area values contributing to the exponentially decaying tail of the area distribution shown in figure \ref{fig:AreaPanel}(a,b). In order to observe the dynamics of single myosin II filaments, tracks including larger aggregates must be removed before analysis.

\subsection{Myosin II Filament tracking}

Tracks of myosin II filament motion were generated by considering the spatial displacement and the change in area of the detected particle between subsequent frames. First, it was checked whether the change in position of a detection between two consecutive frames was suitably small. An estimation of the maximum bound velocity of a moysin II filament of $v_{\textnormal{max}}\sim 0.6\, \mu \textnormal{m}\,\textnormal{s}^{-1}$ suggests filaments can move a maximum distance of $|\Delta \bm{x}|_\textnormal{max} \sim 4 \,\textnormal{pxl}$ between frames. Here, the magnitude of the change in position between frames was limited to $|\Delta \bm{x}| = \sqrt{(\bar{x}_2-\bar{x}_1)^2+(\bar{y}_2-\bar{y}_1)^2} \leq 6 \,\textnormal{pxl}$ for a new detection to be considered part of an existing track (where $\bar{x},\bar{y}$ are the co-ordinates of the centroid of the detected particle, and the subscripts indicate initial ($1$) and final ($2$) positions), to allow for small deviations in motion and shape.

Second, a detection at time $t_2$ was also only added to the track of a detection at time $t_1$ if its area $A(t_2)$ satisfied $A(t_1) - A_{\textnormal{av}} \leq A(t_2) \leq A(t_1) + A_{\textnormal{av}}$, which allowed for the transient overlap of two detections. This can lead to tracks being cut into multiple shorter tracks due to overlap events, which would skew dwell time and total displacement distributions towards lower values. This effect is particularly noticeable in data sets with lots of aggregation or clustering of particles. Based on the intensity and area thresholds, it can be assumed that the detected particles must have been bound for the majority of the exposure time (here 200 ms). However, for the remaining area analysis a detection only associated with tracks of at least two frames duration are included, to further prevent noise affecting the result.

\begin{figure}[t!]
	\includegraphics[width=1.0\linewidth]{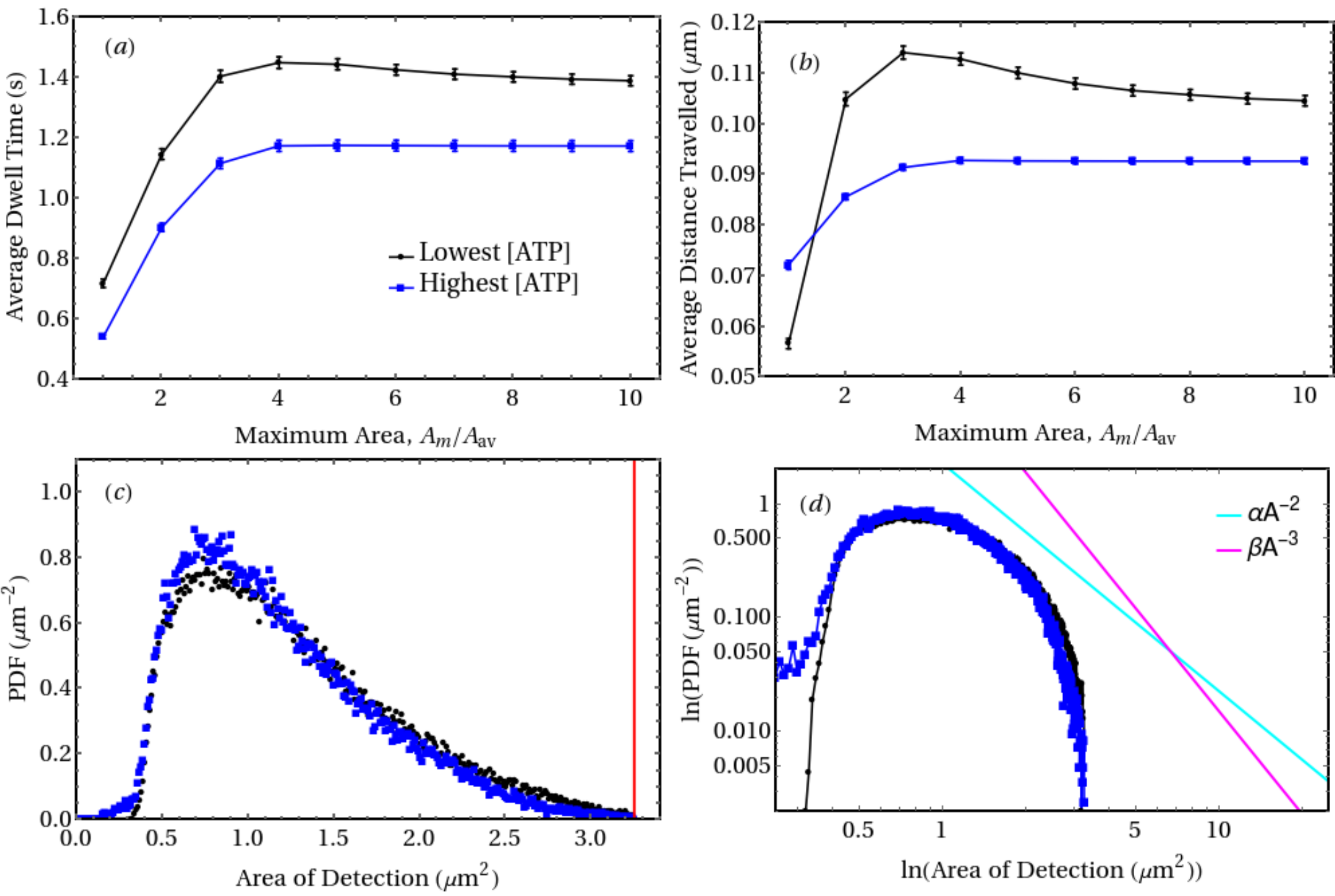}
	\caption{(a) The effect of varying the maximum area cutoff $A_{\textnormal{max}}$ (in multiples of $A_{\textnormal{av}}$) on the average dwell time of the myosin II filaments in the iSCAT videos for the highest and lowest ATP concentration data sets. (b) The effect of varying the maximum area cutoff $A_{\textnormal{max}}$ on the average straight-line distance the myosin II filaments travelled between their initial and final positions in the iSCAT videos. (c) The corrected probability distribution functions of detection areas after the application of the derived minimum and maximum area cutoffs. Red vertical line shows the maximum area cutoff. (d) Log-log plots of the corrected area distributions showing their more favourable decay profiles.}
	\label{fig:TrackingPanel}
\end{figure}

In order to derive an appropriate maximum area threshold that minimises the detection of myosin II filament aggregates, the effect of varying the maximum area (as a multiple of the average area, $A_{\textnormal{av}}$) on average system variables was analysed (figure \ref{fig:TrackingPanel}(a,b)). The maximum area was set to $A_{\textnormal{max}}=4A_{\textnormal{av}}=3.251\,\mu\textnormal{m}^2$ as it maximised the measured average dwell time, while also being larger than the value that maximised the measured average distance travelled. It should be noted that all tracks that contain a point with an area $A>A_{\textnormal{max}}$ have been removed in their entirety from the data to be analysed in order to prevent artificially skewing the dwell time and displacement distributions. In order to measure dwell times and total displacements as accurately as possible, only tracks that ended (by detachment or by leaving the field of view) by the end of the video were included in the analysis. The effect of myosin II filaments leaving the field of view on the measured dwell time was found to be negligible, as less than $4\%$ of tracks ended within $6 \,\textnormal{pxl}$ (the maximum allowed filament displacement between timesteps) of the edge of the video. The final probability distribution function describing the detected areas after applying the above corrections is shown in figure \ref{fig:TrackingPanel}(c,d).

When two myosin II filaments transiently overlapped, the aggregate was always appended to the track with the current longest dwell time in order to probe longer track dynamics. The ratio of tracks that ended within $l=\sqrt{A_{\textnormal{av}}}$ of any point on another track (potentially due to aggregation) was independent of the dwell time of the tracks (results not shown). Similarly, the dwell time distributions and associated characteristic timescales, including the average dwell time, of the tracked myosin II filaments at varying ATP concentrations did not change significantly after the removal of the tracks that ended in an overlap event (results not shown). These results suggest that the overlap events do not introduce a characteristic timescale into the dwell time analysis, and hence can be ignored.

Following particle detection and tracking, analysis is carried out as follows:

\begin{center}
	\begin{tabular}{||c}
		Tracks that contain large displacements that could be due to \\processive particle motion are located (section \ref{sec:Proc}),\\\\
		
		Processive regions of tracks are isolated using bounds on the \\time-correlation of the filament displacements (section \ref{sec:Proc}),\\\\
		
		Filament orientation is analysed to complete the \\parameterisation of each track (section \ref{sec:Angle}).
	\end{tabular}
\end{center}

Following these steps, distributions can be derived for myosin II filament displacements, mean-squared displacements, orientations, mean-squared angular displacements, and dwell times, along entire tracks and separately along regions of purely processive motion. 

\subsection{Defining Processive Regions of Myosin II Filament Tracks \label{sec:Proc}}

Of particular interest in this work is the splitting of tracks into regions of processive, directed motion and diffusive, non-directed motion. The processive, directed motion of the myosin II filaments executed when the filaments are fully bound to the underlying actin network has a much longer persistence length than the diffusive motion exhibited by the filaments when they are only partially bound to the surface. Tracks were defined as being diffusive or processive using the formalism from the work by Jeanneret et al. \cite{Jeanneret2016}; firstly tracks that contain large, directional displacements that cannot be a result of Brownian motion were identified, and then the exact frames corresponding to these displacements were isolated by considering the correlation in the displacements between adjacent timesteps.

A purely diffusive, spherical particle with a time-dependent position $\bm{x}(t)$ and diffusivity $D$ in two-dimensions will always have an average displacement of zero and a mean-squared displacement equal to $\langle \Delta \bm{x}(t)^2 \rangle_D = 4Dt$. Assuming that the myosin II filaments being tracked exhibit non-diffusive motion, this expectation value can be used as a lower bound of the mean-squared displacement required to identify a particle as moving processively. In principle, a filament could exhibit $\langle \Delta \bm{x}(t) \rangle \neq 0$ due to the processive motion of other nearby filaments, but the results of such hydrodynamic effects are unclear and they are ignored in the following derivations.

Due to the anisotropy in the shape of the ellipsoidal myosin II filaments, it is expected that they will exhibit different diffusivities in the directions parallel ($D_a$) and perpendicular ($D_b$) to their orientation (or semi-major axis). This means that, when in the diffusive state, a filament will move with an average diffusivity $D_{avg}=(D_a+D_b)/2$ in the lab frame (which is the quantity measured in this work), but that the individual, directional diffusivities along the $x$ and $y$ axes in the lab frame ($D_x$ and $D_y$ respectively) will only tend to the value $D_{avg}$ at long times \cite{Han2006}. The analytical form of the diffusion tensor for an ellipse shows that $D_x$ and $D_y$ are functions of the initial orientation of the ellipse, and decay monotonically towards the average diffusivity $D_{avg}$ with the timescale \cite{Han2006},

\begin{equation}
\tau_D = \frac{1}{4 D_{\theta}},
\end{equation}

\noindent where the angular diffusivity $D_{\theta}$ implicitly includes information about the average eccentricity of the myosin II filaments. This is by definition the timescale for the diffusion tensor to become isotropic.


Neglecting any single frame detection, the average dwell time of the myosin II filaments at the lowest ATP concentration (corresponding to the longest average dwell time) was $\langle t \rangle = (1.449 \pm 0.019)\,$s after considering $15647$ detected filaments. The angular diffusivity was calculated from the gradient of a linear fit to the first ten points, $2\,$s, of the mean-squared angular displacement data when plotted against time, and for this data set was $D_\theta = (0.077 \pm 0.003)\,\textnormal{rad}^2\,\textnormal{s}^{-1}$. This corresponds to an angular decorrelation time of $\tau_D = (3.27 \pm 0.10)\,$s. The measured angular diffusivity was higher for the data set at the highest ATP concentration, as on average fewer processive regions of track were detected that have high correlation in their orientations. In this case $\tau_D = (1.39 \pm 0.07)\,$s, which is more comparable to the average dwell time for the data set $\langle t \rangle = (1.173 \pm 0.018)\,$s after considering $14016$ detected filaments.


As $\tau_D > \langle t \rangle$ for all sample data sets, individual filament trajectories will maintain a good degree of correlation with their initial direction along their full length, and the diffusivities $D_x$ and $D_y$ along the lab frame axes for individual filament trajectories will be different from the long-term average diffusivity $D_{avg}$. However, their average, $(D_x + D_y) / 2$, provides an unbiased estimate of $D_{avg}$ even for individual trajectories \cite{Han2006}. In this work, we sample a large population of trajectories whose initial orientations are isotropically distributed in the lab frame. In this case, even the single-axis diffusivities $D_x$ and $D_y$ are equal to $D_{avg}$ when the average over the whole set of trajectories is considered. The average diffusivity that we report, $\langle (D_x + D_y) / 2 \rangle$, is averaged over the ensemble of all recorded trajectories, and is an unbiased estimate of $D_{avg}$.

\begin{figure}[t!]
	\includegraphics[width=1.0\linewidth]{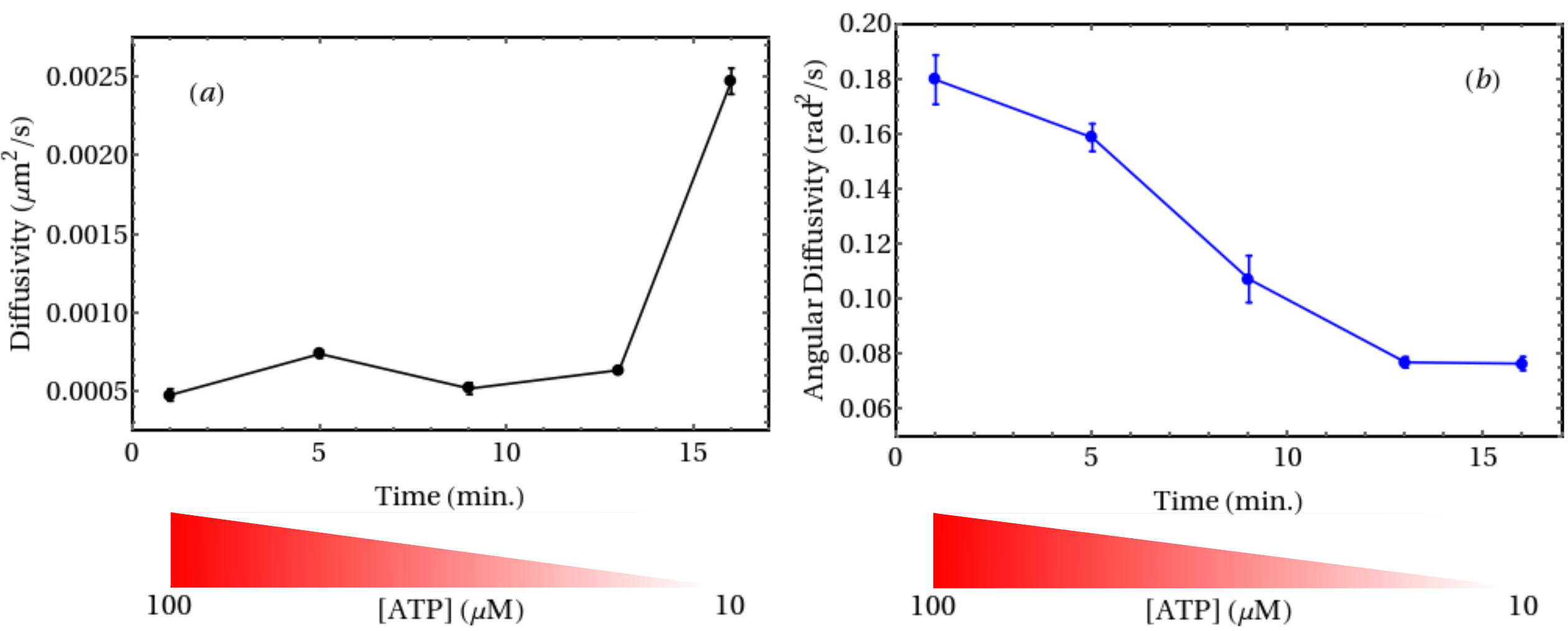}
	\caption{(a) Dependence of myosin II filament (translational) diffusivity on ATP concentration, or the time since the start of the experiment from which the sample data sets were collected. As the ATP concentration decreases, the filaments move further on average from their initial position as a function of time as the fraction of time that filaments spend moving processively increases. This leads to an increased observed average diffusivity. (b) Dependence of filament angular diffusivity on ATP concentration, or the time since the start of the experiment. As the fraction of time that the filaments spend moving processively increases, the width of the distribution describing the difference between their orientation and direction of propagation decreases, resulting in a decrease in the observed angular diffusivity.}
	\label{fig:DiffusivityPanel}
\end{figure}

Tracks with processive regions were isolated by requiring the mean-squared displacement of a myosin II filament to be greater than $16 D_{avg}\Delta t$ ($=4\langle \Delta \bm{x}(t)^2 \rangle_{D_{avg}}$) over a period of $\Delta t =2\,\textnormal{s}$ at some point along its track. This sets the requirement that a filament must have a dwell time of at least $2\,\textnormal{s}$ to be defined as moving processively, but as processive motion has been connected to stronger binding to the actin network and longer dwell times \cite{Mosby2019} we believe this threshold to be reasonable. The value of the diffusivity $D_{avg}$ used to recognise processive tracks was derived from the sample data set with the highest ATP concentration (using the same method as for the value of $D_{\theta}$ above and accounting for the $2$D system), in order to minimise the effects of processive motion on the measurement. It has been assumed that the average filament diffusivity is independent of ATP concentration. The effect of processivity on the measured (translational) diffusivity and angular diffusivity for each of the sample data sets is shown in figure \ref{fig:DiffusivityPanel}. The value of the diffusivity used to recognise processive tracks was $D_{avg}=(0.000\,48 \pm 0.000\,04)\,\mu\textnormal{m}^2\,\textnormal{s}^{-1}$, resulting in a threshold for processivity of $16 D_{avg}\Delta t \sim 0.015\,\mu\textnormal{m}^2$.

\begin{figure}[t!]
	\includegraphics[width=1.0\linewidth]{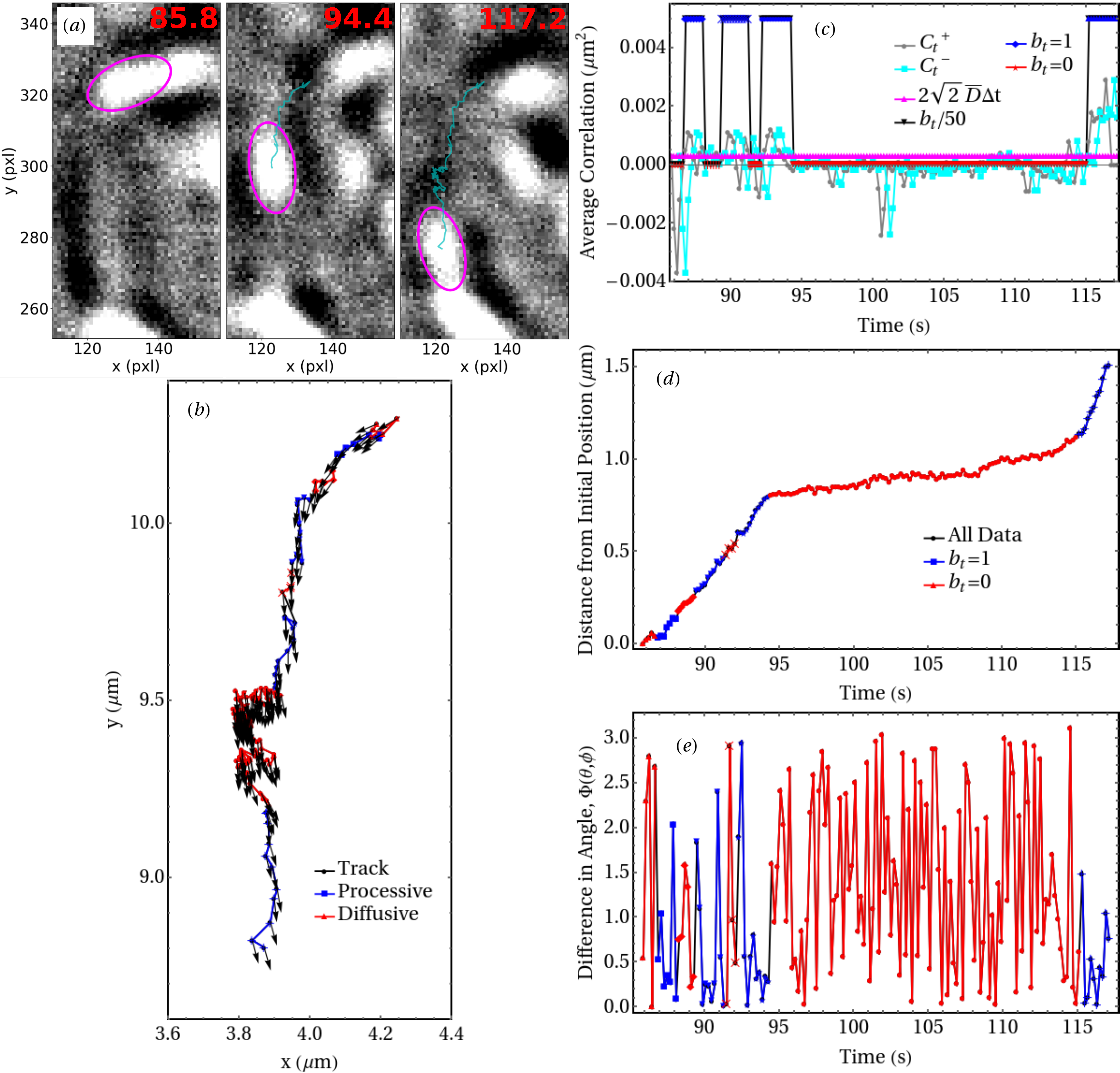}
	\caption{(a) Snapshots from iSCAT video with the lowest ATP concentration showing track of detected myosin II filament (cyan) with elliptical filament shape (magenta), labeled by time since the start of the video in seconds. (b) Track of example myosin II filament split into processive (blue) and diffusive (red) regions. Arrows show filament orientation at each time. (c) Parameters used to define whether a point on the track is processive or diffusive. Regions where either $C_t^+$ or $C_t^-$ (Eqs.~(\ref{eq:DisplacementCorrelation+}~\&~\ref{eq:DisplacementCorrelation-})) are greater than $2\sqrt{2}D_{avg}\Delta t$ for at least $5$ frames are processive, and $b_t=1$. (d) Distance from initial position increases approximately linearly in processive regions, but plateaus in long diffusive region. (e) Difference between orientation angle and direction of propagation (calculated from Eq.~(\ref{eq:AngleDifference})) appears to have more frequent fluctuations with larger amplitudes in the diffusive regions.}
	\label{fig:ProcessivePanel}
\end{figure}

Once a track with a region of large displacement was identified, the points that corresponded to the processive motion were isolated by considering the correlation in displacements between adjacent timesteps \cite{Jeanneret2016}. For a diffusive process this correlation is $\langle \bm{dr}_{t+\Delta t}\cdot \bm{dr}_t \rangle = 0$ (where $\bm{dr}_t = \bm{r}(t+\Delta t) - \bm{r}(t)$), and the variance in this quantity is,

\begin{equation}
\mathrm{Var}(\bm{dr}_{t+\Delta t}\cdot \bm{dr}_t) = 8D_{avg}^2 \Delta t^2,
\end{equation}

\noindent after averaging over all possible initial orientations. This means that the standard deviation in the scalar product $\bm{dr}_{t+\Delta t}\cdot \bm{dr}_t$ is $\sigma(\bm{dr}_{t+\Delta t}\cdot \bm{dr}_t) = 2\sqrt{2}D_{avg}\Delta t$, even for a spatially anisotropic particle. We therefore require that the scalar product between displacements at adjacent timesteps must be greater than the value of $\sigma(\bm{dr}_{t+\Delta t}\cdot \bm{dr}_t)$ for a purely diffusive particle in order for a point to be defined as processive.

In order to minimise the number of false-positive detections of processive points on tracks, the smoothed, average signal,

\begin{equation}
C_{t}^+=\frac{\bm{dr}_{t+2\Delta t}\cdot \bm{dr}_{t+\Delta t} + \bm{dr}_{t+\Delta t}\cdot \bm{dr}_t}{2},
\label{eq:DisplacementCorrelation+}
\end{equation}

\noindent has been calculated as in the work by Jeanneret et al. \cite{Jeanneret2016}, and a myosin II filament is defined as moving processively at time $t$ if $C_{t}^+ > 2\sqrt{2}D_{avg}\Delta t$. Similarly, because the decision of whether or not a filament is moving processively at time $t$ should be invariant under time reversal symmetry, for each point on the track the averaged correlation $C_t$ is calculated for both directions in time. Time inversion generates a new form of Eq.~(\ref{eq:DisplacementCorrelation+}) to be calculated,

\begin{equation}
C_{t}^-=\frac{\bm{dr}_{t-2\Delta t}\cdot \bm{dr}_{t-\Delta t} + \bm{dr}_{t-\Delta t}\cdot \bm{dr}_t}{2},
\label{eq:DisplacementCorrelation-}
\end{equation}

\noindent but the value to compare this to for a purely diffusive particle is invariant under this transformation, $\sigma(\bm{dr}_{t-\Delta t}\cdot \bm{dr}_t) = 2\sqrt{2}D_{avg}\Delta t$. An example of a track that has been split into diffusive and processive regions is shown in figure \ref{fig:ProcessivePanel}.

Using the average correlation in both directions in time to define whether a myosin II filament is moving processively ensures that points on a filament's track at the edge of a processive region are not neglected incorrectly (for example $C_{t}^+$ cannot be defined for the final three points of any track), and that short ($\leq3$ timestep) deviations from processive motion between two processive regions of track are still classed as processive. For example, if a myosin II filament exhibited a single, sharp jump in its propagation direction, and hence processivity, this could be the result of the filament binding to a new actin filament and continuing its processive motion. In this case it would be correct to define this as a single processive region. This method could artificially increase the timescales of observed processive regions by small amounts, but has been used to minimise the impacts of small deviations in filament processivity, such as the artificial splitting of processive regions, and to track edge effects. In this work the added requirement that a point is only defined as processive if it is in a region of at least five similarly defined processive points (over a time of $1\,\textnormal{s}$) is used. A binary signal $b_t$ has been defined such that $b_t=1$ at time $t$ if either $C_{t}^\pm > 2\sqrt{2}D_{avg}\Delta t$, and $b_t=0$ otherwise, as shown in figure \ref{fig:ProcessivePanel}(c).

\subsection{Interpreting Myosin II Filament Orientation \label{sec:Angle}}

The SEP Python package generates the orientation angles of each elliptical particle it detects in an image (above the threshold area and intensity described in section \ref{sec:Detection}) on a $-\pi/2 \leq \theta \leq \pi/2$ domain. The myosin II filaments studied in this work break the expected symmetry along the semi-major axis of an ellipse by having a preferential direction of propagation when bound to actin, and so their orientation must be defined on a $-\pi \leq \theta \leq \pi$ domain instead. Extending this domain further and tracking filament orientation on the domain $-\infty < \theta < \infty$ ensures that no large jumps are observed in filament orientation due to the periodic boundaries of a $-\pi \leq \theta \leq \pi$ domain. This re-parameterisation is especially important for the calculation of angular diffusivity.

\begin{figure}[t!]
	\includegraphics[width=1.0\linewidth]{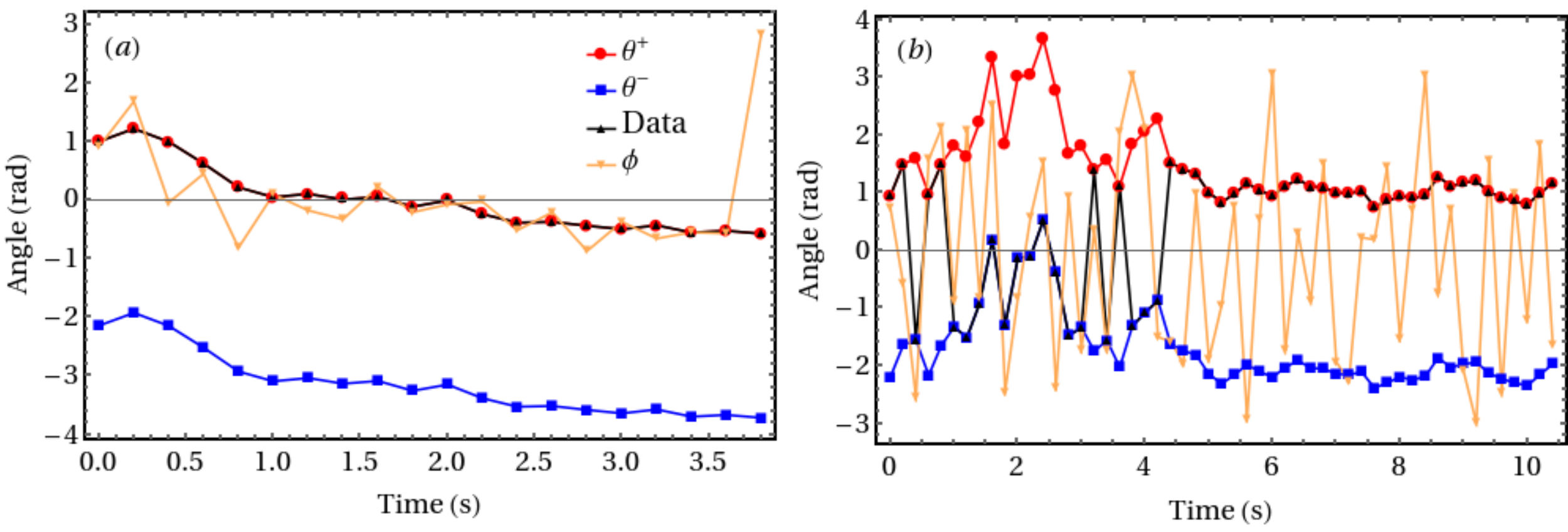}
	\caption{Different angular track properties for single tracks. (a) Purely processive track, propagation direction $\phi$ is aligned with orientation angle $\theta^+$ for majority of track. Black data shows the orientation extracted by the SEP Python package, red and blue data represent possible orientation tracks in angle-space, $\theta^+$ and $\theta^-$ respectively. The $\theta^+$ data maximises the sum in Eq.~(\ref{eq:AngleAverage}). (b) Purely diffusive track, propagation direction $\phi$ varies faster and with greater amplitude than orientation angle $\theta^\pm$. Orientations extracted by the SEP Python package are split amongst both of the possible orientation tracks in angle-space.}
	\label{fig:AnglePanel}
\end{figure}

Of the two possible initial orientation angles (parallel and anti-parallel to the semi-major axis of the elliptical myosin II filament detection), it is assumed that the correct choice will minimise the average (over time) of the difference between the filament's orientation at time $t$, $\theta_t$, and propagation direction at time $t$, $\phi_t=\mathrm{arctan}((\bar{y}_{t+\Delta t}-\bar{y}_{t})/(\bar{x}_{t+\Delta t}-\bar{x}_{t}))$ (where $\bar{x}_t,\bar{y}_t$ are the co-ordinates of the centroid of the detected ellipse at time $t$). This average is calculated over either the processive region(s) of track, or the entire track if the filament exhibits purely diffusive motion. It has been found that filaments moving processively have a smaller average value of $\phi_t - \bar{\theta}_t$ (where $\bar{\theta}_t = (\theta_{t+\Delta t} + \theta_t) / 2$) \cite{Mosby2019}, with examples shown in figure \ref{fig:AnglePanel}, so using processive region(s) to derive the filament's initial orientation is preferred if possible.

Once a detection is added to a track it is assigned the orientation angle (oriented either parallel or anti-parallel to its semi-major axis) that minimises the filament's change in orientation between frames. It has been assumed that a filament cannot rotate by more than $\Delta  \theta_{\textnormal{max}} = \pi / 2$ between frames due to its anisotropic, ellipsoidal shape that causes preferential diffusion along its semi-major axis. This allows the tracking of a filament's orientation on a $-\infty < \theta < \infty$ domain. It has also been assumed that the propagation direction of a filament cannot vary by more than $\Delta \phi_{\textnormal{max}} = \pi$ between frames, so the new propagation direction minimises the change in the directions between frames taking into account the $2\pi$ periodicity of the domain.

Of the two possible orientation tracks in angle-space (that result from the two available initial filament orientations) the selected path is the one that maximises the sum,

\begin{equation}
\Psi(\theta, \phi, T) = \sum_{i=0}^{(T/\Delta t)-1} \mathrm{cos}(\phi_{i\Delta t} - \bar{\theta}_{i\Delta t}),
\label{eq:AngleAverage}
\end{equation}

\noindent where $T$ is the total duration of the iSCAT video being analysed (with corresponding time between frames $\Delta t$). If the filament exhibits processive motion, then the sum in Eq.~(\ref{eq:AngleAverage}) is instead taken from $i=T_s/\Delta t$ to $i=(T_f/\Delta t)-1$, where $T_s$ and $T_f$ are the start and finish times of the processive region(s). The $\mathrm{cos}$ term inside the sum is a maximum when the average filament orientation is aligned with its propagation direction, and is a minimum when the directions are anti-parallel, and implicitly takes into account the $2\pi$ periodicity of the domain. The values $\Psi^\pm(\theta_t^\pm, \phi_t, T)$ corresponding to the two possible orientation paths $\theta_t^\pm$ (labelled as one path will always start with a positive orientation angle and the other a negative one) will always be separated by a factor of $-1$, as,

\begin{equation}
\Psi^\pm(\theta^\pm, \phi, T) = \mathrm{cos}(\pm \pi) \Psi^\mp(\theta^\mp, \phi, T) = - \Psi^\mp(\theta^\mp, \phi, T),
\end{equation}


\noindent so only one path needs to be followed in order to choose the correct initial orientation.

By exploiting the domain of the $\mathrm{arccos}$ function from the Math Python package, the magnitude of the difference between the orientation angle and the direction of propagation can be calculated at each time $t$ as,

\begin{equation}
\Phi(\bar{\theta}_t, \phi_t) = |\mathrm{arccos}( \mathrm{cos}( \phi_{t} - \bar{\theta}_{t} ) )|.
\label{eq:AngleDifference}
\end{equation}

\noindent Using Eq.~(\ref{eq:AngleDifference}), the distance moved by a myosin II filament at each time, $|\Delta \bm{x}_t|=r_t$, can be separated into components parallel, $|\Delta \bm{x}_t|_{\textnormal{para}} = r_t |\mathrm{cos}(\Phi(\bar{\theta}_t, \phi_t))|$, and perpendicular, $|\Delta \bm{x}_t|_{\textnormal{perp}} = r_t |\mathrm{sin}(\Phi(\bar{\theta}_t, \phi_t))|$, to the filament orientation. The error in individual orientation measurements due to image pixelisation and inherent pixel noise can be estimated by using the tracking procedure presented here to observe particles permanently stuck to a surface.

Using the results of this angle tracking, a histogram of $\bar{\theta}_t-\bar{\theta}_0$ can be plotted for each time $t$ after the initial observation of a filament. These results show that the orientation distribution evolves as a Gaussian with width proportional to time $\sim D_\theta t$ (results not shown), and hence that it was correct to use a linear fit to calculate the angular diffusivity from the mean-squared angular displacement data.


\section{Discussion}

The computational methods described here were developed with the aim of analysing the motion of myosin II filaments containing multiple motor protein domains when bound to an underlying actin filament network. Regular point or circular body tracking algorithms could not be used to accurately parameterise the motion of the elongated, ellipsoidal myosin II filaments, and would not be able to provide any information about their angular fluctuations. Splitting tracks into regions of diffusive or processive motion allows us to potentially probe the dynamics of different bound states for the myosin II filaments, which could be extended to the study of different biological systems. Following the development and calibration of the SPT method outlined in this paper, analysis of the full myosin II filament data set has yielded interesting information about the dwell time distribution and spatial dynamics of bound myosin II filaments as a function of the ATP concentration of the system \cite{Mosby2019}. This novel SPT method can be used to study the dynamics of particles at varying length-scales and has potential applications in the fields of fluorescence and light microscopy. Tracking and analysis code is available upon request.


\ack
DVK acknowledges the Kukura lab for enabling the imaging of myosin II filament dynamics on their iSCAT microscopes. DVK thanks the Warwick-Wellcome QBP for funding. LM and MP gratefully acknowledge support from Leverhulme Trust Grant RPG-2016-260.

\appendix
%

\end{document}